\documentclass{article}
\usepackage{spconf,amsmath,graphicx}

\usepackage{color}
\usepackage{multirow}
\usepackage{comment}
\usepackage{subfigure}
\usepackage{colortbl}

\newcommand{\bi}[1]{\ensuremath{\boldsymbol{#1}}}   
\newlength\savedwidth
\newcommand{\wcline}[1]{\noalign{\global\savedwidth\arrayrulewidth\global\arrayrulewidth 1.0pt} \cline{#1}
\noalign{\global\arrayrulewidth\savedwidth}}

\title{Impact of Sound Duration and Inactive Frames\\on Sound Event Detection Performance}
\name{Keisuke Imoto$^{\dagger}$, Sakiko Mishima$^\diamondsuit$, Yumi Arai$^\diamondsuit$, Reishi Kondo$^\diamondsuit$
}
\address{$^{\dagger}$\hspace{1pt}Doshisha University, $^{\diamondsuit}$\hspace{1pt}NEC Corporation}
\begin{document}
\ninept
\maketitle
%
\begin{abstract}
In many methods of sound event detection (SED), a segmented time frame is regarded as one data sample to model training.
The durations of sound events greatly depend on the sound event class, e.g., the sound event ``fan'' has a long duration, whereas the sound event ``mouse clicking'' is instantaneous.
Thus, the difference in the duration between sound event classes results in a serious data imbalance in SED.
Moreover, most sound events tend to occur occasionally; therefore, there are many more inactive time frames of sound events than active frames.
This also causes a severe data imbalance between active and inactive frames.
In this paper, we investigate the impact of sound duration and inactive frames on SED performance by introducing four loss functions, such as simple reweighting loss, inverse frequency loss, asymmetric focal loss, and focal batch Tversky loss.
Then, we provide insights into how we tackle this imbalance problem.
%
\end{abstract}
%
%
\vspace{-3pt}
\section{Introduction}
\label{sec:intro}
\vspace{-3pt}
Sound event detection (SED), in which the types of sound event are identified and their onset and offset in an audio recording are estimated, is one of the principal tasks in environmental sound analysis \cite{Virtanen_Springer2018_01,Imoto_AST2018_01}.
Recently, many works have addressed SED because it plays an important role in realizing various applications using artificial intelligence in sounds, e.g., automatic life logging, machine monitoring, automatic surveillance, media retrieval, and biomonitoring systems \cite{Imoto_INTERSPEECH2013_01,Koizumi_arXiv2020_01,Ntalampiras_ICASSP2009_01,Jin_INTERSPEECH2012_01,Salamon_PLoSOne2016_01,Okamoto_NCSP2020_01}.

\begin{figure}[t]
\centering
\includegraphics[width=0.92\columnwidth]{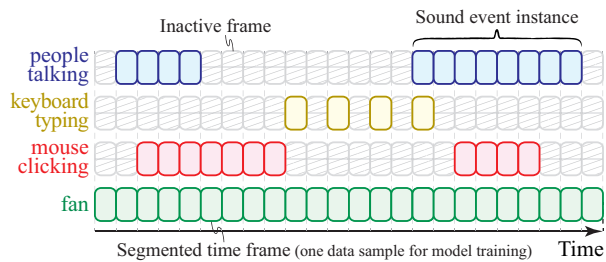}
\vspace{-9pt}
\caption{Examples of active/inactive sound events and number of data samples}
\label{fig:duration}
\end{figure}

\begin{figure*}[t!]
\centering
\includegraphics[width=1.75\columnwidth]{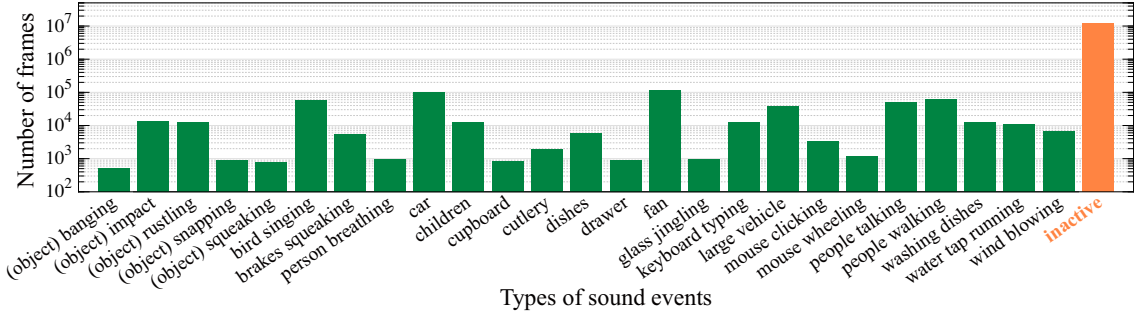}
\vspace{-11pt}
\caption{Numbers of frames of active/inactive sound events in dataset used for evaluation experiments}
\label{fig:num_frame_01}
\vspace{-3pt}
\end{figure*}

For SED, many methods using neural networks, such as a convolutional neural network (CNN) \cite{Hershey_ICASSP2017_01}, recurrent neural network (RNN) \cite{Hayashi_TASLP2017_01}, convolutional recurrent neural network (CRNN) \cite{Cakir_TASLP2017_01}, and Transformer-based neural network \cite{Miyazaki_ICASSP2020_01,Kong_TASLP2020_01}, have been proposed.
In these methods, an audio clip is segmented into short time frames (e.g., 40 ms frames), and each frame is regarded as one data sample for model training and evaluation.
As shown in Fig.~\ref{fig:duration}, sound events vary in duration, and the average frame length of sound events varies highly depending on the sound event class.
Table~\ref{tbl:duration} and Fig.~\ref{fig:num_frame_01} show the average duration of one sound event instance and the total number of frames covered by sound events in development datasets used for evaluation experiments described in Sec. \ref{sec:experiments} (TUT Sound Events 2016, 2017, TUT Acoustic Scenes 2016, and 2017 \cite{Mesaros_EUSIPCO2016_01,Mesaros_DCASE2017_01}), respectively.
In these datasets, the number of frames in the sound event ``mouse clicking,'' which has an average length of 0.15 s, is 1,163, whereas that in the sound event ``fan,'' which has an average length of 29.99 s, is 116,837.
Thus, the difference in the duration between sound events causes a serious data imbalance in SED.
Moreover, Figs.~\ref{fig:duration} and \ref{fig:num_frame_01} show that there are much larger differences in the number of data samples between frames in which sound events are active and inactive.
In the development dataset used for evaluation experiments, the total number of active frames is $5.28 \times 10^{5}$ and that of inactive frames is $1.24 \times 10^{7}$; therefore, this difference also causes a serious imbalance problem between active and inactive data samples.

There are some conventional methods of SED in the case of imbalanced data \cite{Chen_INTERSPEECH2019_01,Wang_ICASSP2020_01,Dinkel_ICASSP2020_01}.
For example, Chen and Jin have proposed a method of detecting rare sound events using data augmentation \cite{Chen_INTERSPEECH2019_01}.
Wang \textit{et al.} have proposed a method of few-shot SED based on metric learning \cite{Wang_ICASSP2020_01}.
Dinkel and Yu have proposed a method of SED using a temporal subsampling method within a CRNN \cite{Dinkel_ICASSP2020_01}.
However, the impact of data imbalance on SED performance caused by the difference in duration between sound event classes and active/inactive frames has not been comprehensively investigated in these works.
Our contributions in this paper are as follows.
\begin{table}[t]
\vspace{-10pt}
\footnotesize
\caption{Average duration of one sound event instance in datasets used for evaluation experiments (TUT Sound Events 2016, 2017, TUT Acoustic Scenes 2016, and 2017 [14, 15])}
\vspace{-7pt}
\label{tbl:duration}
\begin{center}
\begin{tabular}{lrclr}
\wcline{1-2}\wcline{4-5}
&\\[-7pt]
\multicolumn{1}{c}{\textbf{Sound event}} \!\!&\!\!\! \textbf{Duration (s)} \!\!&\!\!\!\!&\multicolumn{1}{c}{\!\! \textbf{Sound event}} \!\!&\!\!\! \textbf{Duration (s)}\\[-1pt]
\wcline{1-2}\wcline{4-5}
&\\[-7.2pt]
(object) banging \!\!\!\!&\!\!\!\! 0.78\ \ \ \ \ \ \ &\!\!\!\!&\!\! drawer \!\!\!\!&\!\!\!\! 0.80\ \ \ \ \ \ \ \\[-1.2pt]
(object) impact \!\!\!\!&\!\!\!\! 0.35\ \ \ \ \ \ \ &\!\!\!\!&\!\! fan \!\!\!\!&\!\!\!\! 29.99\ \ \ \ \ \ \ \\[-1.2pt]
(object) rustling \!\!\!\!&\!\!\!\! 2.24\ \ \ \ \ \ \ &\!\!\!\!&\!\! glass jingling \!\!\!\!&\!\!\!\! 0.80\ \ \ \ \ \ \ \\[-1.2pt]
(object) snapping \!\!\!\!&\!\!\!\! 0.46\ \ \ \ \ \ \ &\!\!\!\!&\!\! keyboard typing \!\!\!\!&\!\!\!\! 0.21\ \ \ \ \ \ \ \\[-1.2pt]
(object) squeaking \!\!\!\!&\!\!\!\! 0.74\ \ \ \ \ \ \ &\!\!\!\!&\!\! large vehicle \!\!\!\!&\!\!\!\! 14.68\ \ \ \ \ \ \ \\[-1.2pt]
bird singing \!\!\!\!&\!\!\!\! 7.63\ \ \ \ \ \ \ &\!\!\!\!&\!\! mouse clicking \!\!\!\!&\!\!\!\! 0.14\ \ \ \ \ \ \ \\[-1.2pt]
brakes squeaking \!\!\!\!&\!\!\!\! 1.65\ \ \ \ \ \ \ &\!\!\!\!&\!\! mouse wheeling \!\!\!\!&\!\!\!\! 0.16\ \ \ \ \ \ \ \\[-1.2pt]
breathing \!\!\!\!&\!\!\!\! 0.43\ \ \ \ \ \ \ &\!\!\!\!&\!\! people talking \!\!\!\!&\!\!\!\! 4.09\ \ \ \ \ \ \ \\[-1.2pt]
car \!\!\!\!&\!\!\!\! 6.88\ \ \ \ \ \ \ &\!\!\!\!&\!\! people walking \!\!\!\!&\!\!\!\! 6.63\ \ \ \ \ \ \ \\[-1.2pt]
children \!\!\!\!&\!\!\!\! 6.87\ \ \ \ \ \ \ &\!\!\!\!&\!\! washing dishes \!\!\!\!&\!\!\!\! 4.15\ \ \ \ \ \ \ \\[-1.2pt]
cupboard \!\!\!\!&\!\!\!\! 0.65\ \ \ \ \ \ \ &\!\!\!\!&\!\! water tap running \!\!\!\!&\!\!\!\! 5.92\ \ \ \ \ \ \ \\[-1.2pt]
cutlery \!\!\!\!&\!\!\!\! 0.74\ \ \ \ \ \ \ &\!\!\!\!&\!\! wind blowing \!\!\!\!&\!\!\!\! 6.09\ \ \ \ \ \ \ \\[0pt]\wcline{4-5}
dishes \!\!\!\!&\!\!\!\! 1.24\ \ \ \ \ \ \ &\!\!\!\!&\!\!&\\[-0.7pt]
\wcline{1-2}
\end{tabular}
\vspace{-13pt}
\end{center}
\end{table}

\vspace{0pt}
\begin{itemize}
  \setlength{\itemsep}{5pt}
  \item We clarify how the data imbalance between sound event classes and/or active and inactive frames impacts the performance of SED.\\[-15.5pt]
  \item We introduce some loss functions into SED tasks, such as asymmetric local loss and focal batch Tversky loss.
\end{itemize}
\vspace{0pt}

The remainder of this paper is constructed as follows.
In Sec. 2, we introduce the conventional SED method using the sigmoid cross-entropy loss function.
In Sec. 3, we introduce some reweighting techniques for the cross-entropy loss function.
In Sec. 4, we determine the impact of the duration of active/inactive sounds on SED performance on the basis of experimental results, and in Sec. 5, we conclude this paper.
%
%
\vspace{-3pt}
\section{Sound Event Detection Using Binary Cross Entropy Loss}
\label{SED}
\vspace{-3pt}
Let us consider a model $f$ and model parameter ${\bi \theta}$.
The goal of SED is to predict sound event labels $\hat{{\bf Z}}$ in an unknown sound by

\vspace{-8pt}
\begin{align}
\hat{{\bf Z}} = I \hspace{-0.6pt} \left[ f({\bf X}, {\bi \theta}) \geq \phi \right] \in \{0,1\}^{N \times M},
\end{align}
%

\noindent where $N$, $M$, ${\bf X}$, $\phi$, and $I$ are the number of time frames in a sound clip, the number of sound event classes, the acoustic feature, the detection threshold, and the indicator function, respectively.
The model parameter ${\bi \theta}$ is preliminarily determined using the training dataset $\mathcal D = \{ ({\bf X}_{1}, {\bf Z}_{1}), ..., ({\bf X}_{l}, {\bf Z}_{l}), ..., $ $({\bf X}_{L}, {\bf Z}_{L}) \hspace{-1pt} \}$.
Here, ${\bf X}_{l}$ is the acoustic feature of the $l^{th}$ sound clip and ${\bf Z}_{l} = \{ {\bf z}_{l,1}, ...$ ${\bf z}_{l,n}, ..., {\bf z}_{l,N}\}$ indicates a sequence of multi--hot vector ${\bf z}_{l,n} \in \{0, 1\}^{M}$ in the $l^{th}$ sound clip over the $M$ sound event class.
For the acoustic feature ${\bf X}_{l}$, the mel-band energy and mel-frequency cepstral coefficients (MFCCs) are often used.
For the model $f$, CNN, CRNN, or a Transformer-based neural network has been applied.
The model parameter ${\bi \theta}$ is estimated using the following binary cross-entropy (BCE) loss function $E_{{\rm BCE}} ({\bi \theta})$ and the backpropagation technique:

\vspace{-11pt}
\begin{align}
{\rm E}_{{\rm BCE}} ({\bi \theta}) &= - \sum^{N}_{n=1} {\Big \{} {\bf z}_{n} \log ( {\bf y}_{n} ) + (1-{\bf z}_{n}) \log (1-{\bf y}_{n} ) {\Big \}} \nonumber\\[-3pt]
&\hspace{-20pt} = - \!\! \sum^{N\!, M}_{n,m=1} \!\! {\Big \{} z_{n,m} \log ( y_{n,m} ) + (1 - z_{n,m}) \log ( 1 - y_{n,m} ) {\Big \}},\nonumber\\[-6pt]
\label{eq:conv_loss}
\end{align}
\vspace{-12pt}

\noindent where $y_{n,m}$ are the prediction of sound event $m$ in time frame $n$.
$z_{n,m}$ is the target label in time frame $n$ and is 1 if acoustic event $m$ is active in time frame $n$ and 0 otherwise.
Note that the sound clip index $l$ is omitted to simplify the equation.
${\rm E}_{{\rm BCE}} ({\bi \theta})$ is calculated by summing the binary cross-entropy over all time frames and sound event classes.
Since the duration of sound events varies highly depending on the event class, the model parameter estimation using the BCE loss leads to the data imbalance between sound event classes.
Moreover, as shown in Figs.~\ref{fig:duration} and \ref{fig:num_frame_01}, because the number of inactive frames of sound events is much larger than that of active frames, the model parameter estimation tends to be overwhelmed by the inactive frames.
Consequently, active frames tend to be ignored in the model training.
%
\vspace{-3pt}
\section{Loss Function Considering Data Imbalance}
\label{Method}
\vspace{-5pt}
In this work, we apply four loss functions that can control the contribution to model training of short/long sound events and active/inactive frames.
%
\vspace{-9pt}
\subsection{Simple Reweighting Loss (SRL)}
\label{ssec:SRL}
\vspace{-5pt}
To investigate the impact of a very large number of inactive frames on the SED performance, we consider the following simple reweighting loss (SRL):

\vspace{-9pt}
\begin{align}
{\rm E}_{\rm SRL} ({\bi \theta}) &= - \! \sum^{N\!, M}_{n,m=1} \! {\Big \{} \alpha \hspace{0.3pt} z_{n,m} \log ( y_{n,m} ) \nonumber\\[-4pt]
&\hspace{22pt} + \beta (1 - z_{n,m}) \log ( 1 - y_{n,m} ) {\Big \}},
\label{eq:SRL}
\end{align}
\vspace{-9pt}

\noindent where $\alpha \in [0, \infty)$ and $\beta \in [0, \infty)$ are the reweighting factors.
In this work, we set $\alpha$ as 1.0.
%

\vspace{-7pt}
\subsection{Inverse Frequency Loss (IFL)}
\label{ssec:IFL}
\vspace{-4pt}
To investigate the impact of data imbalance between sound event classes, we also consider the following reweighted loss on the basis of the inverse frequency of sound events (IFL):

\vspace{-9pt}
\begin{align}
{\rm E}_{{\rm IFL}} ({\bi \theta}) &= - \! \sum^{N\!, M}_{n,m=1} \! \bigg\{ \Big( \frac{C}{N_{m} + C} \Big)^{\!\gamma} z_{n,m} \log ( y_{n,m} ) \nonumber\\[-5pt]
&\hspace{25pt} + (1 - z_{n,m}) \log ( 1 - y_{n,m} ) {\bigg \}},
\label{eq:IFL}
\end{align}
\vspace{-7pt}

\noindent where $\gamma \in [0, \infty)$, $N_{m}$, and $C$ are the wighting factor, the number of frames of a sound event $m$ in a training batch, and a constant, respectively.
The IFL can reweight the contribution of each sound event for model training in accordance with the frequency, and enable more robust model training with the imbalanced dataset.
%
%
\vspace{-7pt}
\subsection{Asymmetric Focal Loss (AFL)}
\label{ssec:AFL}
\vspace{-4pt}
Many long-duration sound events (e.g., ``fan'' and ``car'') and inactive durations are stationary, that is, they have less variation in their acoustic features, and their model training is easy.
On the other hand, several sound events of short duration (e.g., ``object impact'' and ``keyboard typing'') have more than one audio pattern, such as attack, decay, and release parts.
To control the training weight of the sound event model in accordance with the ease/difficulty of model training, the use of focal loss has been proposed \cite{Lin_ICCV2017_01,Noh_Sensors2020_01}.
In this paper, we newly introduce the following asymmetric focal loss (AFL), which enables the control of the focusing factor of active and inactive frames separately.

\vspace{-10pt}
\begin{align}
{\rm E}_{{\rm AFL}} ({\bi \theta}) &= - \! \sum^{N\!, M}_{n, m=1} {\Big \{} (1 - y_{n,m})^{\gamma} \hspace{1pt} z_{n,m} \log ( y_{n,m} ) \nonumber\\[-4pt]
&\hspace{25pt} + (y_{n,m})^{\zeta} \hspace{1pt} (1 - z_{n,m}) \log ( 1 - y_{n,m} ) {\Big \}}\nonumber\\[-20pt]
\label{eq:AFL}
\end{align}
\vspace{-11pt}

\noindent Here, $\gamma$ and $\zeta$ are the weighting parameters that control the focusing weight of active and inactive frames, respectively.
When we set large values for the weighting parameters $\gamma$ and $\zeta$, the loss of active and inactive frames is down-weighted depending on the prediction error.
%
\vspace{-15pt}
\subsection{Focal Batch Tversky Loss (FBTL)}
\label{ssec:TLDL}
\vspace{-3pt}
To address the data imbalance between active and inactive frames, we also introduce the focal batch Tversky loss (FBTL) ${\rm E}_{{\rm FBTL}} ({\bi \theta})$ into the SED task, which is an extended version of the focal dice loss \cite{Li_ACL2020_01} and the Tversky loss \cite{Salehi_MLMI2017_01,Kodym_GCPR2018_01}.
The dice loss and Tversky loss directly train the model to maximize the F-score, which does not consider the true-negative samples.
That is, these losses can also prevent the model training from being overwhelmed by the easy-negative frames.
In this paper, we introduce the idea of focal loss into the batch Tversky loss \cite{Salehi_MLMI2017_01,Kodym_GCPR2018_01}, and apply the following FBTL into the SED task:

\vspace{-11pt}
\begin{align}
&\hspace*{-3.5pt} {\rm E}_{{\rm FBTL}} ({\bi \theta}) = \nonumber\\[-1pt]
&\hspace{3pt} 1 \! - \! \frac{\sum_{l,n,m=1}^{B\!, N\!, M} (1 \! - \! y_{l,n,m}^{(1)})^{\gamma} y_{l,n,m}^{(1)} z_{l,n,m}^{(1)} + \eta}{\sum_{l\hspace{-0.4pt},n\hspace{-0.4pt},m=1}^{B\!, N\!, M} \alpha (1 \! - \! y_{l\hspace{-0.4pt},n\hspace{-0.4pt},m}^{(1)})^{\gamma} y_{l\hspace{-0.4pt},n\hspace{-0.4pt},m}^{(1)} \! + \! \sum_{l\hspace{-0.4pt},n\hspace{-0.4pt},m=1}^{B\!, N\!, M} \beta z_{l\hspace{-0.4pt},n\hspace{-0.4pt},m}^{(1)}  \! + \! \eta},\!
\label{eq:FBTL}
\end{align}
\vspace{-2pt}

\noindent where $y_{l,n,m}^{(1)}$ and $z_{l,n,m}^{(1)}$ are the prediction and sound event label for the active frame, respectively.
$\alpha \in [0, 1.0]$ and $\beta \in [0, 1.0]$ are the tradeoff parameters between false negative and false positive samples, where $\alpha + \beta = 1.0$.
$B$ and $\eta$ is a number of sound clip in each batch and a smoothing parameter, respectively.
\begin{table}[t]
\vspace{-10pt}
\footnotesize
\caption{Experimental conditions}
\label{tbl:parameter}
\centering
\begin{tabular}{ll}
\wcline{1-2}
&\\[-7pt]
\!Length of sound clip \!\!&\!\! 10 s\!\!\\
\cline{1-2}
\!&\\[-8pt]
\!Network for CNN-BiGRU\!\!&\!\! 3 CNN $+$ 1 BiGRU $+$ 1 dense layer\!\!\\[0pt]
\!\# channels of CNN layers \!\!&\!\! 128, 128, 128\!\!\\[0pt]
\!Filter size \!\!&\!\! 3$\times$3, 3$\times$3, 3$\times$3\!\!\\[0pt]
\!Pooling size \!\!&\!\! 1$\times$8, 1$\times$4, 1$\times$2 (max pooling)\!\!\\[0pt]
\!\# units in GRU layer \!\!&\!\! 32\!\!\\[0pt]
\!\# units in fully conn. layer \!\!&\!\! 32\!\!\\[0pt]
\cline{1-2}
\!&\\[-8pt]
\!\multirow{2}{*}{Network for Transformer}\!\!&\!\! 3 CNN $+$ 2 Transformer encoder\!\!\\[-1.5pt]
\!\!\!&layers $+$ 2 dense layers\!\!\\[1pt]
\!\# attention heads\!\!&\!\! 32\!\!\\[0pt]
\cline{1-2}
\!&\\[-8pt]
\!Activation function\!\!&\!\! Leaky ReLU\!\!\\[0pt]
\!Optimizer\!\!&\!\! RAdam \cite{Liu_ICLR2020_01}\!\!\\[0pt]
\!Detection threshold\!\!&\!\! 0.5\!\!\\[0pt]
\!Constant number $C$\!\!&\!\! 500\!\!\\
\!Smoothing parameter $\eta$\!\!&\!\! 1.0\!\!\\
\wcline{1-2}
\end{tabular}
\vspace{-4pt}
\end{table}
\begin{table*}[t!]
\footnotesize
\vspace{0pt}
\caption{Average SED performance for various loss functions and networks}
\vspace{1pt}
\centering
\begin{tabular}{lcccc}
\wcline{1-5}
&&\\[-8pt]
\!\cellcolor[rgb]{0.95,0.95,0.95} &\cellcolor[rgb]{0.95,0.95,0.95} \!\!&\!\!\cellcolor[rgb]{0.95,0.95,0.95} \!\!&\!\! \textbf{\cellcolor[rgb]{0.95,0.95,0.95} Micro-} \!\!&\!\! \textbf{\cellcolor[rgb]{0.95,0.95,0.95} Macro-} \\[-1pt]
\multicolumn{1}{c}{\multirow{-1.9}{*}{\textbf{\cellcolor[rgb]{0.95,0.95,0.95} Method}}}\!\!&\multicolumn{1}{c}{\multirow{-1.9}{*}{\textbf{\cellcolor[rgb]{0.95,0.95,0.95} \!\!Micro-Fscore\!\!}}}&\multicolumn{1}{c}{\multirow{-1.9}{*}{\textbf{\cellcolor[rgb]{0.95,0.95,0.95} \!\!Macro-Fscore\!\!}}}&\!\!\cellcolor[rgb]{0.95,0.95,0.95} \textbf{ROC AUC}\!\!&\!\!\cellcolor[rgb]{0.95,0.95,0.95} \textbf{ROC AUC} \!\!\\[-1pt]
\wcline{1-5}
&&\\[-8pt]
\textbf{[Conventional methods]} \\[0pt]
\ \ \ \ \ \ \ CNN-BiGRU w/ BCE loss (\textbf{Baseline}) &40.10\% &  7.39\% & 89.15\% & 65.85\% \\[0pt]
\ \ \ \ \ \ \ CNN-BiGRU w/ $\alpha$ min-max subsampling \& BCE loss &44.12\% & 9.35\% & 90.27\% & 67.55\% \\[0pt]
\ \ \ \ \ \ \ CNN-BiGRU w/ batch dice loss &45.06\% & 9.79\% & 86.99\% & 63.89\% \\[0pt]
\ \ \ \ \ \ \ MTL of SED \& SAD w/ BCE loss &43.35\% & 8.64\% & 91.40\% & 70.97\% \\[0pt]
\ \ \ \ \ \ \ Transformer w/ BCE loss &45.15\% & 9.27\% & 90.32\% & 66.64\% \\[0pt]
\cline{1-5}
&&\\[-8pt]
\textbf{[Loss reweighting between active and inactive frames]} \\[0pt]
\ \ \ \ \ \ \ CNN-BiGRU w/ simple reweighting loss ($\beta = 0.3535$)&46.44\% & 10.34\% & 91.07\% & 69.31\% \\[0pt]
\ \ \ \ \ \ \ CNN-BiGRU w/ asymmetric focal loss ($\gamma \! = \! 0.0$, $\zeta \! = \! 1.414$) &47.78\% & 10.65\% & 92.35\% & 76.18\% \\[0pt]
\ \ \ \ \ \ \ CNN-BiGRU w/ focal batch Tversky loss ($\alpha \! = \! 0.6, \beta \! = \! 0.4, \gamma \! = \! 0.001$) &46.97\% & 10.28\% & 87.95\% & 65.08\% \\[0pt]
\cline{1-5}
&&\\[-8pt]
\textbf{[Loss reweighting between sound event classes]} \\[0pt]
\ \ \ \ \ \ \ CNN-BiGRU w/ inverse frequency loss ($C=500$) &41.89\% & 7.57\% & 89.89\% & 66.46\% \\[0pt]
\ \ \ \ \ \ \ CNN-BiGRU w/ asymmetric focal loss ($\gamma \! = \! 0.125$, $\zeta \! = \! 0.0$) &42.33\% & 8.13\% & 91.08\% & 70.46\% \\[0pt]
\cline{1-5}
&&\\[-8pt]
\textbf{[Loss reweighting both between event classes and between active/inactive frames]}\!\!\!\\[0pt]
\ \ \ \ \ \ \ CNN-BiGRU w/ asymmetric focal loss ($\gamma \! = \! 0.0625$, $\zeta \! = \! 1.0$) &48.29\% & 10.46\% & 92.62\% & 77.03\% \\[0pt]
\ \ \ \ \ \ \ Transformer w/ asymmetric focal loss ($\gamma \! = \! 0.0625$, $\zeta \! = \! 1.0$) &\textbf{49.14\%} & \textbf{11.11\%} & \textbf{92.74\%} & \textbf{77.49\%} \\[0pt]
\wcline{1-5}
\end{tabular}
\vspace{5pt}
\label{tbl:comparison01}
\end{table*}
%
%
%
\vspace{-3pt}
\section{Experiments}
\label{sec:experiments}
\vspace{-4pt}
\subsection{Experimental Conditions}
\label{ssec:conditions}
\vspace{-3pt}
We evaluated the impact of data imbalance between sound event classes and active/inactive frames on the event detection performance using various SED networks and loss functions.
For the evaluation, we used the dataset composed of parts of TUT Sound Events 2016, 2017, TUT Acoustic Scenes 2016, and 2017 \cite{Mesaros_EUSIPCO2016_01,Mesaros_DCASE2017_01}.
From these datasets, we selected a total of 266 min of sound clips (development set, 192 min; evaluation set, 74 min) including the 25 types of sound event listed in Fig.~\ref{fig:num_frame_01}.
The details of the dataset can be found in \cite{Imoto_dataset2019_01}.
Note that the datasets were recorded not for the detection task of rare sounds but for the analysis of real-life sounds; thus, the analysis of seriously imbalanced data is a general problem in SED.

\begin{figure}[t]
\centering
\includegraphics[width=0.92\columnwidth]{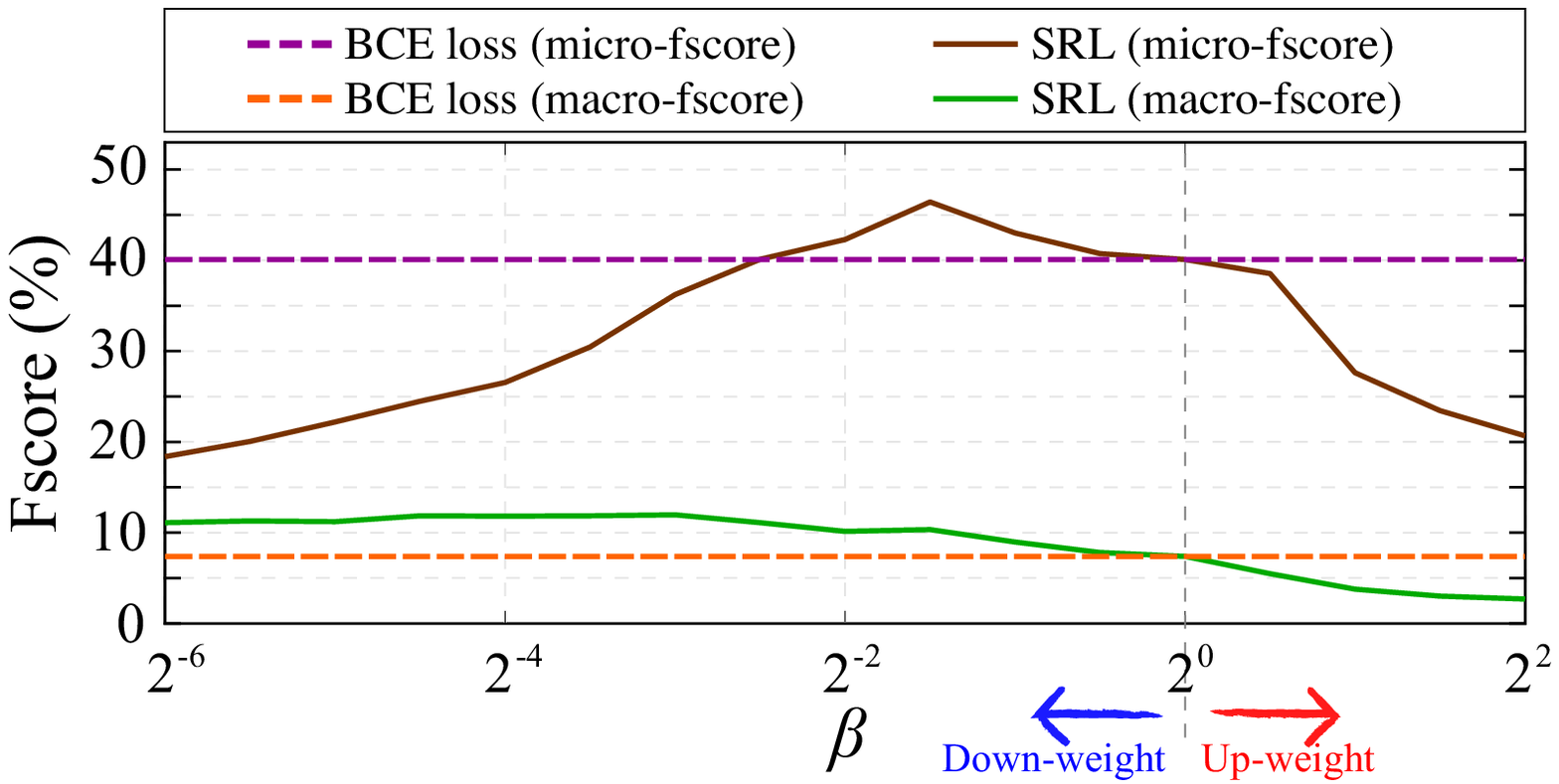}
\vspace{-12pt}
\caption{Average Fscores for SRL and BCE with various weighting factors $\beta$}
\label{fig:result01}
\vspace{8pt}
%
%
\centering
\includegraphics[width=0.92\columnwidth]{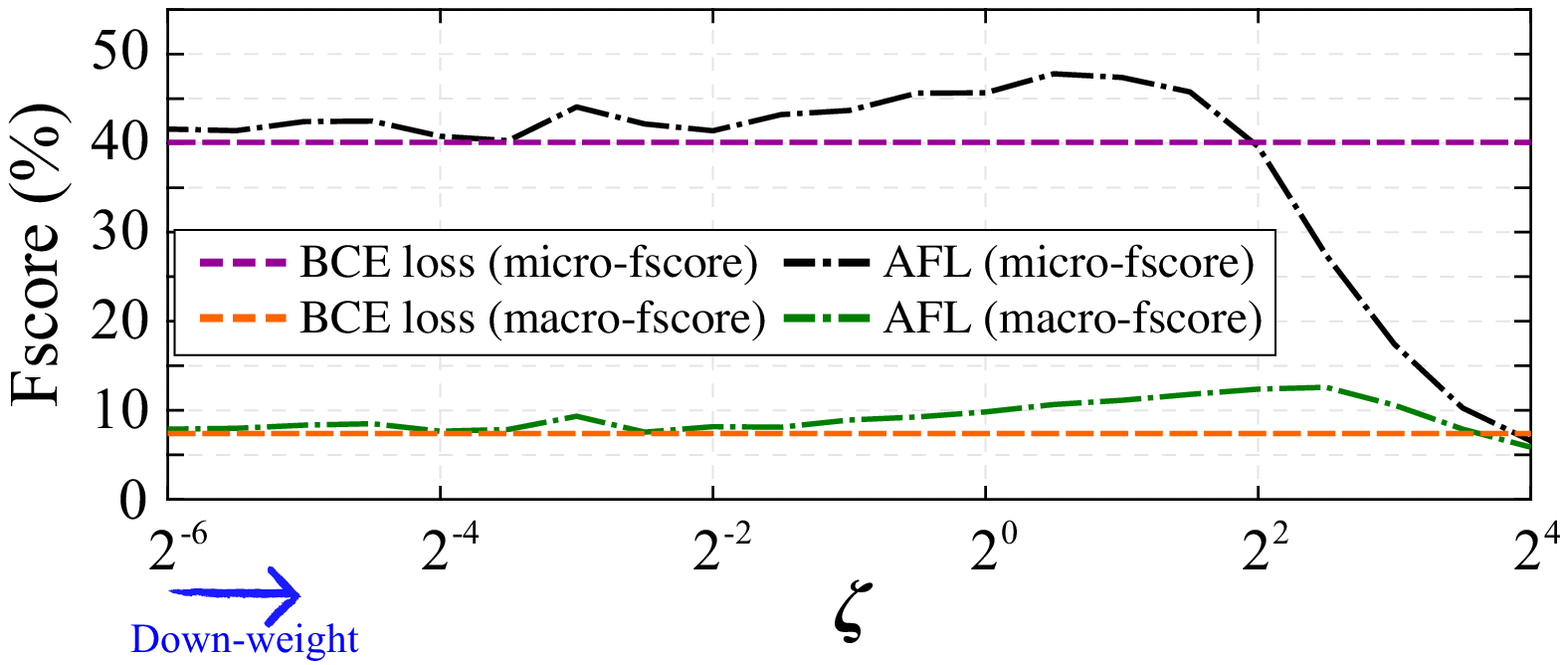}
\vspace{-12pt}
\caption{Average Fscores for AFL and BCE with various weighting factors $\zeta$}
\label{fig:result02}
\vspace{-5pt}
\end{figure}

For the acoustic features, we extracted the 64-dimensional log mel-band energy at a sampling rate of 16 kHz, which was calculated every 40 ms with a 20 ms hop size.
As the baseline network, we used CNN-BiGRU, which is widely used as the baseline system of SED, such as DCASE2018 challenge task 4 \cite{Serizel_DCASE2018_01}.
For each method, we conducted the evaluation experiment 10 times with random initial values for model parameters.
The performance of SED was evaluated using the frame-based macro- and micro-Fscores.
Other experimental conditions are listed in Table~\ref{tbl:parameter}.
%
%
\vspace{-5pt}
\subsection{Experimental Results}
\label{ssec:results}
\vspace{-2pt}
\subsubsection{Impact of inactive frames on SED performance}
\vspace{-3pt}
Figures \ref{fig:result01} and \ref{fig:result02} show the average macro- and micro-Fscores with BCE loss, SRL, and AFL ($\gamma = 0.0$) for various reweighting factors.
In this experiment, we used CNN-BiGRU as the network architecture.
The results show that when the loss of inactive frames is down-weighted, both macro- and micro-Fscores tend to improve.
This indicates that the inactive frames overwhelm model training, which leads to active sound events being ignored in model training.\\[-5pt]
%
\vspace{-12pt}
\subsubsection{Impact of imbalance between event classes on SED performance}
\begin{figure}[t!]
\vspace{-7pt}
\centering
\includegraphics[width=0.92\columnwidth]{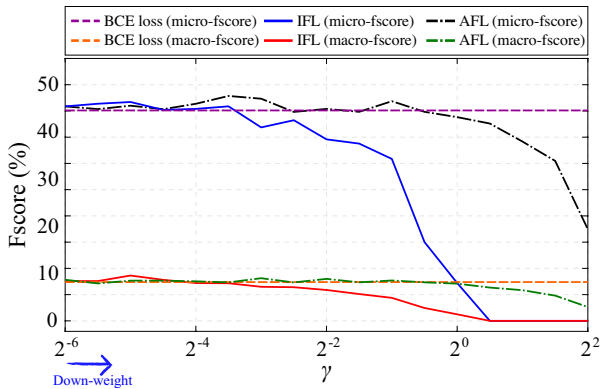}
\vspace{-9pt}
\caption{Average fscores for BCE loss, IFL, and AFL with various weighting factors $\gamma$. For AFL, we set $\zeta = 0.0$.}
\label{fig:result03}
\vspace{-2pt}
\end{figure}

\vspace{-3pt}
Figure \ref{fig:result03} shows the average macro- and micro-Fscores with BCE loss, IFL, and AFL ($\zeta = 0.0$) for various reweighting factors $\gamma$.
The results show that even when the losses are reweighted to be more balanced between sound event classes, both macro- and micro-Fscores do not improve much.
This implies that the data imbalance between sound event classes have less impact on SED performance than the imbalance between active and inactive frames.
Thus, in SED, the data imbalance between active and inactive frames is a more serious problem and needs to be addressed preferentially.
Moreover, when both the imbalance between sound event classes and that between active and inactive frames are reweighted as shown at the bottom of Table~\ref{tbl:comparison01}, both macro- and micro-Fscores are more improved than by reweighting of the imbalance either between event classes or between active and inactive frames.

%

\vspace{-8pt}
\subsubsection{Comparison with conventional methods}
\vspace{-2pt}
We have compared the SED performance of our methods with those of other conventional methods, such as SED using a $\alpha$ min-max subsampling method within CRNN \cite{Dinkel_ICASSP2020_01}, the batch dice loss-based SED \cite{Kodym_GCPR2018_01,Park_DCASE2020_01}, multitask learning of SED and sound activity detection \cite{Pankajakshan_WASPAA2019_01}, and Transformer-based SED \cite{Miyazaki_ICASSP2020_01,Kong_TASLP2020_01}.
As the Transformer-based SED, we used three CNN layers with the same structure as the CNN-BiGRU, followed by two Transformer encoder layers and two dense layers.
Table~\ref{tbl:comparison01} shows the micro- and macro-Fscores, micro-ROC AUC, and macro-ROC AUC.
The results show that down-weighting the loss of inactive frames using SRL, AFL, and FBTL improves both the Fscore and ROC AUC score to a great extent than conventional methods.
Surprisingly, even the SRL with CNN-BiGRU outperforms the BCE loss with the Transformer-based SED system, which is the state-of-the-art technique in SED.
However, FBTL does not improve micro- or macro-ROC AUC.
This is because FBTL does not consider the true-negative samples during model training; therefore, the false-positive rate tends to be worse than those in the conventional BCE loss-based methods.
The SED performance is finally improved by 9.04 and 3.72 percentage points in macro- and micro-Fscores  by reweighting both types of imbalance using AFL and applying the Transformer network compared with the baseline system.
%
\vspace{-8pt}
\subsubsection{Detailed detection results for each sound event}
\vspace{-2pt}
To investigate the SED performance in detail, we show the Fscores for selected sound events in Table~\ref{tab:result04}.
In many sound events, IFL and AFL ($\gamma = 0.125$, $\zeta = 0.0$) do not markedly improve the SED performance, whereas AFL ($\gamma = 0.0625$, $\zeta = 1.0$) achieves the best results.
This finding also supports the notion that inactive frames have an adverse effect on SED performance, and reweighting the imbalance between sound event classes and between active and inactive frames improves the performance considerably.
However, even when we apply the reweighting methods, sound events that have a quite small number of frames such as ``drawer,'' are not well detected.
These sound events should be considered in future experiments.

\begin{table}[t!]
\vspace{-20pt}
\caption{Average Fscore for each sound event}
\label{tab:result04}
\hspace*{-4pt}
\footnotesize
\centering
\renewcommand{\arraystretch}{1.0}
\begin{tabular}{lcccccc}
\wcline{1-6}\\
\cellcolor[rgb]{0.95,0.95,0.95} &\cellcolor[rgb]{0.95,0.95,0.95} &\cellcolor[rgb]{0.95,0.95,0.95} &\cellcolor[rgb]{0.95,0.95,0.95} &\cellcolor[rgb]{0.95,0.95,0.95} &\cellcolor[rgb]{0.95,0.95,0.95} &\\[-17pt]
\!\!\cellcolor[rgb]{0.95,0.95,0.95}\!\!\!&\!\!\!{\bf \cellcolor[rgb]{0.95,0.95,0.95} bird}\!\!\!&\!\!\!\cellcolor[rgb]{0.95,0.95,0.95} \!\!\!&\!\!\!\!\cellcolor[rgb]{0.95,0.95,0.95} \!\!\!&\!\!\!\!{\bf \cellcolor[rgb]{0.95,0.95,0.95} washing}\!\!\!&\!\!\!\!{\bf \cellcolor[rgb]{0.95,0.95,0.95} water tap}\!\!\\
\multicolumn{1}{c}{\multirow{-1.9}{*}{\!\!\! \cellcolor[rgb]{0.95,0.95,0.95} \bf Method}}&\!\!{\bf \cellcolor[rgb]{0.95,0.95,0.95} singing}\!\!\!&\!\!\!\multirow{-1.9}{*}{\!\!\!\bf \cellcolor[rgb]{0.95,0.95,0.95} car}\!\!\!&\!\!\!\multirow{-1.9}{*}{\bf \cellcolor[rgb]{0.95,0.95,0.95} drawer}\!\!\!&\!\!\!{\bf \cellcolor[rgb]{0.95,0.95,0.95} dishes}\!\!\!&\!\!\!\!{\bf \cellcolor[rgb]{0.95,0.95,0.95} running}\!\!\\
\wcline{1-6}\\[-7pt]
\!\!CNN-BiGRU w/ BCE\!\!\!&\!\!\!17.79\%\!\!\!\!&\!\!\!\!43.85\%\!\!\!\!\!&\!\!\!0.00\%\!\!\!\!&\!\!\!0.41\%\!\!\!\!&\!\!\!\!43.23\%\!\!\!\\
\!\!\hspace{-2pt}\cellcolor[rgb]{0.95,0.95,0.95} CNN-BiGRU w/ SRL\!\!\!&\!\!\!\cellcolor[rgb]{0.95,0.95,0.95} 32.69\%\!\!\!&\!\!\!\!\cellcolor[rgb]{0.95,0.95,0.95} 49.09\%\!\!\!\!&\!\!\!\cellcolor[rgb]{0.95,0.95,0.95} 0.00\%\!\!\!&\!\!\!\cellcolor[rgb]{0.95,0.95,0.95} 5.09\%\!\!\!&\!\!\!\!\cellcolor[rgb]{0.95,0.95,0.95} 69.37\%\!\!\\
\!\!CNN-BiGRU w/ IFL\!\!\!&\!\!\!19.28\%\!\!\!\!&\!\!\!\!44.42\%\!\!\!\!\!&\!\!\!\!0.00\%\!\!\!\!&\!\!\!0.39\%\!\!\!\!&\!\!\!\!32.76\%\!\!\!\\
\!\!\hspace{-2pt}\cellcolor[rgb]{0.95,0.95,0.95} CNN-BiGRU w/ FBTL\!\!\!&\!\!\!{\bf \cellcolor[rgb]{0.95,0.95,0.95} 45.35}\%\!\!\!&\!\!\!\!\cellcolor[rgb]{0.95,0.95,0.95} 48.88\%\!\!\!\!&\!\!\!\cellcolor[rgb]{0.95,0.95,0.95} 0.00\%\!\!\!&\!\!\!\cellcolor[rgb]{0.95,0.95,0.95} 3.73\%\!\!\!&\!\!\!\!\cellcolor[rgb]{0.95,0.95,0.95} 60.31\%\!\!\\
\!\!CNN-BiGRU w/ AFL\!\!\!&\!\!\!\multirow{2}{*}{34.50\%}\!\!\!&\!\!\!\!\multirow{2}{*}{47.96\%}\!\!\!\!&\!\!\!\multirow{2}{*}{0.00\%}\!\!\!&\!\!\!\multirow{2}{*}{3.94\%}\!\!\!&\!\!\!\!\multirow{2}{*}{73.90\%}\!\!\\[-2pt]
\!\!($\gamma = 0.0$, $\zeta = 1.414$)\!\!\!&\!\!\!\!\!\!&\!\!\!&\!\!\!\!\!\!&\!\!\!\!\!\!&\!\!\!\!\!\\[1pt]
\!\!\hspace{-2pt}\cellcolor[rgb]{0.95,0.95,0.95} CNN-BiGRU w/ AFL\!\!\!&\!\!\!\cellcolor[rgb]{0.95,0.95,0.95} \!\!\!&\!\!\!\!\cellcolor[rgb]{0.95,0.95,0.95} \!\!\!\!&\!\!\!\cellcolor[rgb]{0.95,0.95,0.95} \!\!\!&\!\!\!\cellcolor[rgb]{0.95,0.95,0.95} \!\!\!&\!\!\!\!\cellcolor[rgb]{0.95,0.95,0.95} \!\!\\[-2pt]
\!\!\hspace{-2pt}\cellcolor[rgb]{0.95,0.95,0.95} ($\gamma = 0.125$, $\zeta = 0.0$)\!\!\!&\!\!\!\multirow{-1.6}{*}{\cellcolor[rgb]{0.95,0.95,0.95} 20.15\%}\!\!\!&\!\!\!\!\multirow{-1.6}{*}{\cellcolor[rgb]{0.95,0.95,0.95} \cellcolor[rgb]{0.95,0.95,0.95} 43.35\%}\!\!\!\!&\!\!\!\multirow{-1.6}{*}{\cellcolor[rgb]{0.95,0.95,0.95} 0.00\%}\!\!\!&\!\!\!\multirow{-1.6}{*}{0\cellcolor[rgb]{0.95,0.95,0.95} .74\%}\!\!\!&\!\!\!\!\multirow{-1.6}{*}{\cellcolor[rgb]{0.95,0.95,0.95} 40.29\%}\!\!\\[0pt]
\!\!CNN-BiGRU w/ AFL\!\!\!&\!\!\!\multirow{2}{*}{28.08\%}\!\!\!&\!\!\!\!\multirow{2}{*}{45.71\%}\!\!\!\!&\!\!\!\multirow{2}{*}{0.00\%}\!\!\!&\!\!\!\multirow{2}{*}{10.55\%}\!\!\!&\!\!\!\!\multirow{2}{*}{74.64\%}\!\!\\[-2pt]
\!\!($\gamma = 0.0625$, $\zeta = 1.0$)\!\!\!\!\!&\!\!\!\!\!\!&\!\!\!&\!\!\!\!\!\!&\!\!\!\!\!\!&\!\!\!\!\!\\[1pt]
\!\!\hspace{-2pt}\cellcolor[rgb]{0.95,0.95,0.95} Transformer w/ AFL\!\!\!&\!\!\!\cellcolor[rgb]{0.95,0.95,0.95} \!\!\!&\!\!\!\!\cellcolor[rgb]{0.95,0.95,0.95} \!\!\!\!&\!\!\!\cellcolor[rgb]{0.95,0.95,0.95} \!\!\!&\!\!\!\cellcolor[rgb]{0.95,0.95,0.95} \!\!\!&\!\!\!\!\cellcolor[rgb]{0.95,0.95,0.95} \!\!\\[-2pt]
\!\!\cellcolor[rgb]{0.95,0.95,0.95} ($\gamma = 0.0625$, $\zeta = 1.0$)\!\!\!\!\!&\!\!\!\multirow{-1.6}{*}{\cellcolor[rgb]{0.95,0.95,0.95} 18.40\%}\!\!\!&\!\!\!\!\multirow{-1.6}{*}{{\bf \cellcolor[rgb]{0.95,0.95,0.95} 49.95\%}}\!\!\!\!&\!\!\!\multirow{-1.6}{*}{{\bf \cellcolor[rgb]{0.95,0.95,0.95} 0.03\%}}\!\!\!&\!\!\!\multirow{-1.6}{*}{{\bf \cellcolor[rgb]{0.95,0.95,0.95} 26.71\%}}\!\!\!&\!\!\!\!\multirow{-1.6}{*}{{\bf \cellcolor[rgb]{0.95,0.95,0.95} 79.31\%}}\!\!\\[0pt]
\wcline{1-6}
\end{tabular}
\vspace{0pt}
\end{table}
%
%
%
\vspace{-5pt}
\section{Conclusion}
\label{sec:conclusion}
\vspace{-5pt}
In this work, we studied the impact of the data imbalance between sound event classes and between active and inactive frames on SED performance.
To investigate their impact, we introduced the four loss functions SRL, IFL, AFL, and FBTL, which can reweight the losses and relieve the imbalance of contribution of model training.
The experimental results showed that the inactive frames tend to overwhelm model training; consequently, the trained model tends to ignore active sound events.
On the other hand, the results also show that the data imbalance between sound event classes have less impact on SED performance than the imbalance between active and inactive frames.
To avoid this negative impact on SED performance, the SED methods using asymmetric focal loss and focal batch Tversky loss are effective and considerably improve the SED performance.
%
\vspace{-5pt}
\section{Acknowledgement}
\label{sec:ack}
\vspace{-5pt}
This work was supported by JSPS KAKENHI Grant Number JP19K20304.
\newpage
\vspace{0pt}
\bibliographystyle{IEEEtran}
\bibliography{IEEEabrv,ICASSP2021refs,KeisukeImoto10}

\begin{thebibliography}{10}
\providecommand{\url}[1]{#1}
\def\UrlFont{\rmfamily}
\providecommand{\newblock}{\relax}
\providecommand{\bibinfo}[2]{#2}
\providecommand\BIBentrySTDinterwordspacing{\spaceskip=0pt\relax}
\providecommand\BIBentryALTinterwordstretchfactor{4}
\providecommand\BIBentryALTinterwordspacing{\spaceskip=\fontdimen2\font plus
\BIBentryALTinterwordstretchfactor\fontdimen3\font minus
  \fontdimen4\font\relax}
\providecommand\BIBforeignlanguage[2]{{%
\expandafter\ifx\csname l@#1\endcsname\relax
\typeout{** WARNING: IEEEtran.bst: No hyphenation pattern has been}%
\typeout{** loaded for the language `#1'. Using the pattern for}%
\typeout{** the default language instead.}%
\else
\language=\csname l@#1\endcsname
\fi
#2}}

\bibitem{Virtanen_Springer2018_01}
T.~Virtanen, M.~Plumbley, and D.~Ellis, Eds., \emph{Computational Analysis of
  Sound Scenes and Events}.\hskip 1em plus 0.5em minus 0.4em\relax Springer,
  2017.

\bibitem{Imoto_AST2018_01}
K.~Imoto, ``Introduction to acoustic event and scene analysis,''
  \emph{Acoustical Science and Technology}, vol.~39, no.~3, pp. 182--188, 2018.

\bibitem{Imoto_INTERSPEECH2013_01}
K.~Imoto, S.~Shimauchi, H.~Uematsu, and H.~Ohmuro, ``User activity estimation
  method based on probabilistic generative model of acoustic event sequence
  with user activity and its subordinate categories,'' \emph{Proc.
  INTERSPEECH}, 2013.

\bibitem{Koizumi_arXiv2020_01}
Y.~Koizumi, Y.~Kawaguchi, K.~Imoto, T.~Nakamura, Y.~Nikaido, R.~Tanabe,
  H.~Purohit, K.~Suefusa, T.~Endo, M.~Yasuda, and N.~Harada, ``{DCASE}2020
  challenge task2: Unsupervised anomalous sound detection for machine condition
  monitoring,'' \emph{arXiv, arXiv:2006.05822}, pp. 1--5, 2020.

\bibitem{Ntalampiras_ICASSP2009_01}
S.~Ntalampiras, I.~Potamitis, and N.~Fakotakis, ``On acoustic surveillance of
  hazardous situations,'' \emph{Proc. {IEEE} International Conference on
  Acoustics, Speech and Signal Processing {\rm (}ICASSP{\rm )}}, pp. 165--168,
  2009.

\bibitem{Jin_INTERSPEECH2012_01}
Q.~Jin, P.~F. Schulam, S.~Rawat, S.~Burger, D.~Ding, and F.~Metze,
  ``Event-based video retrieval using audio,'' \emph{Proc. INTERSPEECH}, pp.
  2085--2088, 2012.

\bibitem{Salamon_PLoSOne2016_01}
J.~Salamon, J.~P. Bello, A.~Farnsworth, M.~Robbins, S.~Keen, H.~Klinck, and
  S.~Kelling, ``Towards the automatic classification of avian flight calls for
  bioacoustic monitoring,'' \emph{PLoS One}, vol.~11, no.~11, 2016.

\bibitem{Okamoto_NCSP2020_01}
Y.~Okamoto, K.~Imoto, N.~Tsukahara, K.~Sueda, R.~Yamanishi, and Y.~Yamashita,
  ``Crow call detection using gated convolutional recurrent neural network,''
  \emph{Proc. {RISP} International Workshop on Nonlinear Circuits,
  Communications and Signal Processing {\rm (}NCSP{\rm )}}, pp. 171--174, 2020.

\bibitem{Hershey_ICASSP2017_01}
S.~Hershey, S.~Chaudhuri, D.~P.~W. Ellis, J.~F. Gemmeke, A.~Jansen, R.~C.
  Moore, M.~Plakal, D.~Platt, R.~A. Saurous, B.~Seybold, M.~Slaney, R.~J.
  Weiss, and K.~Wilson, ``{CNN} architectures for large-scale audio
  classification,'' \emph{Proc. {IEEE} International Conference on Acoustics,
  Speech and Signal Processing {\rm (}ICASSP{\rm )}}, pp. 131--135, 2017.

\bibitem{Hayashi_TASLP2017_01}
T.~Hayashi, S.~Watanabe, T.~Toda, T.~Hori, J.~L. Roux, and K.~Takeda,
  ``Duration-controlled {LSTM} for polyphonic sound event detection,''
  \emph{{IEEE/ACM} Trans. Audio Speech Lang. Process.}, vol.~25, no.~11, pp.
  2059--2070, 2017.

\bibitem{Cakir_TASLP2017_01}
E.~\c{C}akir, G.~Parascandolo, T.~Heittola, H.~Huttunen, and T.~Virtanen,
  ``Convolutional recurrent neural networks for polyphonic sound event
  detection,'' \emph{{IEEE/ACM} Trans. Audio Speech Lang. Process.}, vol.~25,
  no.~6, pp. 1291--1303, 2017.

\bibitem{Miyazaki_ICASSP2020_01}
K.~Miyazaki, T.~Komatsu, T.~Hayashi, S.~Watanabe, T.~Toda, and K.~Takeda,
  ``Weakly-supervised sound event detection with self-attention,'' \emph{Proc.
  {IEEE} International Conference on Acoustics, Speech and Signal Processing
  {\rm (}ICASSP{\rm )}}, pp. 66--70, 2020.

\bibitem{Kong_TASLP2020_01}
Q.~Kong, Y.~Xu, W.~Wang, and M.~D. Plumbley, ``Sound event detection of weakly
  labelled data with {CNN}-{T}ransformer and automatic threshold
  optimization,'' \emph{{IEEE/ACM} Trans. Audio Speech Lang. Process.},
  vol.~28, pp. 2450--2460, 2020.

\bibitem{Mesaros_EUSIPCO2016_01}
A.~Mesaros, T.~Heittola, and T.~Virtanen, ``{TUT} database for acoustic scene
  classification and sound event detection,'' \emph{Proc. European Signal
  Processing Conference {\rm (}EUSIPCO{\rm )}}, pp. 1128--1132, 2016.

\bibitem{Mesaros_DCASE2017_01}
A.~Mesaros, T.~Heittola, A.~Diment, B.~Elizalde, A.~Shah, B.~Raj, and
  T.~Virtanen, ``{DCASE} 2017 challenge setup: Tasks, datasets and baseline
  system,'' \emph{Proc. Workshop on Detection and Classification of Acoustic
  Scenes and Events {\rm (}DCASE{\rm )}}, pp. 85--92, 2017.

\bibitem{Chen_INTERSPEECH2019_01}
Y.~Chen and H.~Jin, ``Rare sound event detection using deep learning and data
  augmentation,'' \emph{Proc. {INTERSPEECH}}, pp. 619--623, 2019.

\bibitem{Wang_ICASSP2020_01}
Y.~Wang, J.~Salamon, N.~J. Bryan, and J.~P. Bello, ``Few-shot sound event
  detection,'' \emph{Proc. {IEEE} International Conference on Acoustics, Speech
  and Signal Processing {\rm (}ICASSP{\rm )}}, pp. 81--85, 2020.

\bibitem{Dinkel_ICASSP2020_01}
H.~Dinkel and K.~Yu, ``Duration robust weakly supervised sound event
  detection,'' \emph{Proc. {IEEE} International Conference on Acoustics, Speech
  and Signal Processing {\rm (}ICASSP{\rm )}}, pp. 311--315, 2020.

\bibitem{Lin_ICCV2017_01}
T.~Y. Lin, P.~Goyal, R.~Girshick, K.~He, and P.~Doll\'{a}r, ``Focal loss for
  dense object detection,'' \emph{Proc. {IEEE} International Conference on
  Computer Vision {\rm (}ICCV{\rm )}}, pp. 2980--2988, 2017.

\bibitem{Noh_Sensors2020_01}
K.~Noh and J.~H. Chang, ``Joint optimization of deep neural network-based
  dereverberation and beamforming for sound event detection in multi-channel
  environments,'' \emph{Sensors}, vol.~20, no.~7, pp. 1--13, 2020.

\bibitem{Li_ACL2020_01}
X.~Li, X.~Sun, Y.~Meng, J.~Liang, F.~Wu, and J.~Li, ``Dice loss for
  data-imbalanced {NLP} tasks,'' \emph{Proc. 58th Annual Meeting of the
  Association for Computational Linguistics {\rm (}ACL{\rm )}}, pp. 465--476,
  2020.

\bibitem{Salehi_MLMI2017_01}
S.~S.~M. Salehi, D.~Erdogmus, and A.~Gholipour, ``{T}versky loss function for
  image segmentation using {3D} fully convolutional deep networks,''
  \emph{Proc. International Workshop on Machine Learning in Medical Imaging
  {\rm (}MLMI{\rm )}}, pp. 379--387, 2017.

\bibitem{Kodym_GCPR2018_01}
O.~Kodym, M.~Spanel, and A.~Herout, ``Segmentation of head and neck organs at
  risk using {CNN} with batch dice loss,'' \emph{German Conference in Pattern
  Recognition {\rm (}GCPR{\rm )}}, pp. 105--114, 2018.

\bibitem{Liu_ICLR2020_01}
L.~Liu, H.~Jiang, P.~He, W.~Chen, X.~Liu, J.~Gao, and J.~Han, ``On the variance
  of the adaptive learning rate and beyond,'' \emph{Proc. International
  Conference on Learning Representations {\rm (}ICLR{\rm )}}, pp. 1--13, 2020.

\bibitem{Imoto_dataset2019_01}
\url{https://www.ksuke.net/dataset}

\bibitem{Serizel_DCASE2018_01}
R.~Serizel, N.~Turpault, H.~Eghbal-Zadeh, and A.~P. Shah., ``Large-scale weakly
  labeled semi-supervised sound event detection in domestic environments,''
  \emph{Proc. Workshop on Detection and Classification of Acoustic Scenes and
  Events {\rm (}DCASE{\rm )}}, pp. 19--23, 2018.

\bibitem{Park_DCASE2020_01}
S.~Park, S.~Suh, and Y.~Jeong, ``Sound event localization and detection with
  various loss functions,'' \emph{Technical report of task 3 of {DCASE}
  Challenge 2020}, pp. 1--5, 2020.

\bibitem{Pankajakshan_WASPAA2019_01}
A.~Pankajakshan, H.~L. Bear, and E.~Benetos, ``Polyphonic sound event and sound
  activity detection: A multi-task approach,'' \emph{Proc. {IEEE} Workshop on
  Applications of Signal Processing to Audio and Acoustics {\rm (}WASPAA{\rm
  )}}, pp. 323--327, 2019.

\end{thebibliography}
%
%
\end{document}